\documentclass[]{aa}
\usepackage{graphicx}
\usepackage{natbib}

\def\hh{H$_2$}
\def\dd{D$_2$}
\def\hhhp{H$_3^+$}
\def\hhdp{H$_2$D$^+$}
\def\ddhp{D$_2$H$^+$}
\def\dddp{D$_3^+$}
\def\nnhp{N$_2$H$^+$}

\def\ammo{NH$_3$}
\def\nddd{ND$_3$}
\def\dammo{ND$_3$}
\def\nhhd{NH$_2$D}
\def\nddh{ND$_2$H}
\def\nddd{ND$_3$}

\newcommand   {\about} {\mbox{$\sim$}}
\newcommand   {\arcm}  {\mbox{$^\prime$}}
\newcommand   {\arcs}  {\mbox{$^{\prime\prime}$}}
\newcommand   {\dcop}  {\mbox{DCO$^+$}}
\newcommand   {\pscm}  {\mbox{cm$^{-2}$}}

\newcommand {\arcmper} {\mbox{\rlap{\hbox{\hbox{.}}}\hbox{$^{\prime}$}}}

\def\kkms{K~km~s$^{-1}$}
\def\kms{km~s$^{-1}$}

\def\scm{cm$^{-2}$}
\def\ccm{cm$^{-3}$}

\def\txc{$T_{\rm ex}$}

\def\tmb{$T_{\rm mb}$}
\def\tsys{$T_{\rm sys}$}

\def\apj{ApJ}
\def\apjl{ApJ}
\def\aap{A\&A}

\begin{document}
\title{Interstellar deuterated ammonia: From \ammo\ to \dammo}
\titlerunning{From \ammo\ to \dammo.}

\author{E.~Roueff$^1$, D.C.~Lis$^2$, F.F.S. van der Tak$^3$, M.~Gerin$^4$\and P.F.~Goldsmith$^5$}
\authorrunning{Roueff et al.}
\offprints{E.~Roueff}
\institute{
$^1$ Laboratoire Univers et Th\'{e}orie, UMR 8102 du CNRS,
Observatoire de Paris, Section de Meudon, Place Jules Janssen, 92195
Meudon, France\\
$^2$ Downs Laboratory of Physics 320-47, California Institute of
Technology, Pasadena, CA 91125, USA\\
$^3$ Max-Planck-Institut f{\"u}r Radioastronomie, Auf dem
H{\"u}gel 69, D-53121~Bonn, Germany \\
$^4$ Laboratoire d'Etude du Rayonnement et de la Mati\`{e}re en
Astrophysique, UMR 8112 du CNRS, Observatoire de Paris and Ecole
Normale Sup\'{e}rieure, 24 Rue Lhomond, 75231 Paris Cedex 05, France\\
$^5$ Department of Astronomy and National Astronomy and Ionosphere Center, 
Cornell University, Ithaca, NY 14853, USA\\
}

\date{Received / Accepted}

\abstract{We use spectra and maps of \nhhd, \nddh, and \dammo, obtained with
  the CSO, IRAM 30m and Arecibo telescopes, to study
  deuteration processes in dense cores. 
  The data include the first detection of the hyperfine structure in
  \nddh. The emission of \nhhd\ and \nddd\ does not seem to peak at
  the positions of the embedded protostars, but instead at offset
  positions, where outflow interactions may occur.
  A constant ammonia fractionation ratio in star-forming regions is
 generally assumed to be consistent with an origin on dust grains.
 However, in the pre-stellar cores studied here, the fractionation
 varies significantly when going from \ammo\ to \dammo. We present a
 steady state model of the gas-phase chemistry for these sources,
 which includes passive depletion onto dust grains and multiply
 saturated deuterated species up to five deuterium atoms (e.g.
 CD$_5^+$). The observed column density ratios of all four ammonia
 isotopologues are reproduced within a factor of 3 for a gas
 temperature of 10~K. We also predict that deuterium fractionation
 remains significant at temperatures up to about 20~K. ND and NHD,
 which have rotational transitions in the submillimeter domain are
 predicted to be abundant.

\keywords{ISM: molecules -- Molecular processes -- Stars:
  circumstellar matter -- Stars: formation}
}

\maketitle

\section{Introduction}
\label{sec:intro}

In recent years, the chemistry of dense cores has been revised with
the almost simultaneous discovery of multiply deuterated molecules
(D$_2$CO: \citealt{turn90}, \citealt{cecc98}; ND$_2$H:
\citealt{roueff00}; ND$_3$: \citealt{lis02:nd3}, \citealt{vdtak02};
CHD$_2$OH, CD$_3$OH: \citealt{parise02,parise04}) on the one hand, and
of very large depletions of CO on the other hand \citep{bacmann02}. In
fact, the two processes are closely related since the condensation of
CO and other abundant gas phase molecules favors deuterium
fractionation, by decreasing the destruction rate of deuterated
molecular ions \citep{bacmann03}. As multiply deuterated molecules are
found in dense cores prior to star formation (as in Barnard~1 and
LDN~1689N), or in very young protostars (as in NGC~1333 IRAS~4A), they
can be used for studying and identifying the earliest stages of star
formation. While an analysis of the line profiles gives clues about
the gas dynamics in the envelope surrounding the protostar, or in the
dense core, the molecular abundance and especially the fractionation
provides information on the physical conditions (temperature,
depletion) and on the chemistry. It is therefore of interest to
improve our understanding of deuterium chemistry. For this purpose,
accurate fractionation ratios are needed for multiply deuterated
molecules.

Two main paths are invoked for understanding the large deuterium
fractionation observed in dense cores. The first is based on gas phase
chemistry and invokes the ion-molecule deuterium exchange reactions
taking place at low temperatures, as first proposed by
\citet{watson74}. In this framework, deuterated molecules are built in
successive reactions starting with either H$_2$D$^+$ (and also
D$_2$H$^+$ and D$_3^+$), CH$_2$D$^+$ or C$_2$HD$^+$
(\citealt{roberts00}, \citealt{roberts03}, \citealt{gerlich02}). The
confirmed detections of H$_2$D$^+$ in the envelopes of disks
encompassing the protostars NGC~1333 IRAS~4A
and IRAS~16293-2422 (\citealt{stark99,stark04}) and in the pre-stellar core
LDN~1544 \citep{caselli03}, as well as the recent detection of
D$_2$H$^+$ in LDN~1689N \citep{vastel04}, have provided strong
observational support for this theory. The second path for forming
deuterated molecules is based on grain chemistry (\citealt{tiel83}).
As hydrogenation on grain mantles has been proposed to explain the
large abundances of solid phase CH$_3$OH (\citealt{dart99},
\citealt{gibb00}, \citealt{pontoppidan03}), the analogous process with
deuterium would build deuterated molecules on grains. Because in dense
cores, the gas phase abundance ratio of atomic D/H is probably orders
of magnitude larger than the elemental abundance ratio of $1.5 \times
10^{-5}$, the surface chemistry leads also to large deuterium
fractionation, which scales as the (gas-phase) D/H ratio\footnote{This
paper uses full names (e.g., deuterium) to denote elements, and
symbols (e.g., D) to denote atomic species.}. While the surface
processes seem interesting, there is no secure confirmation of the
role of dust mantles for deuterium fractionation, as neither HDO nor
deuterated methanol has yet been detected in the solid phase
(\citealt{dart03,par03}).

These two paths may lead to different fractionation ratios for
multiply deuterated molecules \citep{rodg01}, because the formation of
deuterated molecules involves largely different processes and chemical
reactions. For gas phase processes, the fractionation ratio at each
stage results from competing chemical reactions (fractionation
reactions substituting deuterium atoms for hydrogen atoms, destruction
reactions with other molecules and dissociative recombination
reactions removing the fractionation, etc.). The fractionation ratio
must be calculated with a comprehensive chemical model and is expected
to be different for the different stages of deuterium fractionation of
a given species, since the specific reactions are different. In a
simple model of grain surface chemistry, the abundance ratios of
various deuterated species are expected to be similar, since each such
ratio is proportional to the gas phase D/H abundance ratio. For all
bonds with a heavy atom, the probability of bonding with a H or D atom
scales with the deuterium/hydrogen ratio. All bonds behave independently in this
process.
 
Four different isotopologues of ammonia, including fully deuterated
ammonia, ND$_3$, as well as NH$_2$D and ND$_2$H, can be observed in
dense cores. The ammonia data can thus provide the information
required for understanding the formation of multiply deuterated
molecules in dense cores. This paper presents new data on deuterated
ammonia in a small sample of dense cores, some of which are forming
stars. The observations are presented in Sect.~\ref{sec:obs}, and the
results described in Sect.~\ref{sec:res}. Sect.~\ref{sec:models}
presents comprehensive gas phase chemical models aimed at
understanding the roles of gas density, depletion, and temperature in
the fractionation of ammonia. We summarize our results in
Sect.~\ref{sec:con}.

\section{Observations}
\label{sec:obs}

The rotational transitions and frequencies of the various
isotopologues of ammonia that have been observed are shown in
Table~\ref{obs:lines}. The frequencies refer to the strongest
hyperfine component. For Einstein $A-$values and statistical weights,
see \citet{tine00} (\nhhd), \citet{roueff00} (\nddh) and
\citet{vdtak02} and \citet{lis02:nd3} (\dammo).

\begin{table}[htbp]
  \caption{\label{obs:lines} Parameters of observed rotational transitions}
  \begin{center}
  \begin{tabular}{lcrc}
  \hline \hline

Species   & Transition    & $E_u$ & $\nu$  \\
          &               &  K    & MHz    \\
\hline
\nhhd   & $1_{11}-1_{01}$ & 20.7  & \phantom{0}85926.3 \\
\nhhd   & $1_{11}-1_{01}$ & 21.3  & 110153.6 \\
\nddh   & $1_{10}-1_{01}$ & 18.7  & 110812.9 \\
\nddh   & $1_{10}-1_{01}$ & 18.4  & 110896.7 \\
\nddh   & $2_{20}-2_{11}$ & 56.8  & 206971.9 \\
\nhhd   & $3_{22}-3_{12}$ & 120.  & 216562.6 \\
\nddd   & $1_{01}-0_{00}$ & 14.9  & 309909.7 \\ 
\hline
\end{tabular}
\end{center}
\end{table}

\subsection{Millimeter-wave observations}
\label{sec:iram}

Observations of \nhhd\ and \nddh\ lines near 110, and 220~GHz were
carried out at the 30-m telescope of the Institut de Radio Astronomie
Millim\'etrique (IRAM) on Pico Veleta, Spain, in August 2002. We also
use additional 86~GHz observations performed in November 2001 and
September 2002. The front ends were the facility receivers A100, B100,
A230 and B230, and the back end was the Versatile Spectral and
Polarimetric Array (VESPA) autocorrelator. During night time, the
weather was good, with $\approx$4~mm of precipitable water vapour and
\tsys$\approx$140~K at 110~GHz, while during daytime, the water column
and \tsys\ roughly doubled. At 110~GHz, the telescope has a FWHM beam
size of $22''$. The pointing of the
telescope was stable to within 2 arcsec. 
Data were calibrated onto a \tmb\ scale by multiplying
by 1.25, the ratio of the forward coupling efficiency to the main beam
efficiency.

Sources were selected from \citet{shah01} to have high \nhhd\ column
densities, but the positions used here are from \citet{jijina99} for
Orion and Monoceros and from \citet{mcmullin00} for Serpens. Table
\ref{tab:pos} lists source positions and LSR velocities. Only upper
limits were obtained toward NGC~2264G and CB~17.

\begin{table}[htbp]
  \caption{\label{tab:pos} Positions and velocities of the
observed sources.}
  \begin{center}
  \begin{tabular}{lrrr}
\hline \hline
\noalign{\smallskip}
Source & R.A.    & Dec.  & V$_{LSR}$ \\
       & J2000 & J2000 & km s$^{-1}$ \\
\hline
\noalign{\smallskip}
NGC~1333        & 03:29:10.3    &   +31:13:32.2  &  7.0 \\
Barnard~1       & 03:33:20.8    &   +31:07:34.8  &  6.8 \\
LDN~1544C       & 05:04:16.6    &   +25:10:47.8  &  7.4 \\
OMC~2           & 05:35:26.3    &  --05:09:49.4  & 11.2 \\
HH~1            & 05:36:17.3    &  --06:46:08.2  &  9.2 \\
LDN~1630        & 05:46:07.4    &  --00:13:41.8  & 10.0 \\
NGC~2264C      & 06:41:11.6    &   +09:29:25.1  &  8.0 \\
LDN~134N        & 15:54:08.5    &  --02:52:01.0  &  2.4 \\
LDN~1689N       & 16:32:29.5    &  --24:28:52.6  &  3.8 \\
LDN~483         & 18:17:29.7    &  --04:39:38.3  &  5.5 \\
S~68N           & 18:29:47.5    &   +01:16:51.4  &  8.5 \\
NGC~2264G       & 06:41:11.0 & +09:56:00.8 & 4.6 \\
CB~17      & 04:04:38.0 & +56:56:11.0 & -4.6 \\
\hline
\end{tabular}
\end{center}
\end{table}

To locate the `deuterium peak' in each source, maps were made of the
\nhhd\ $1_{11}\to1_{01}$ emission near 110 GHz. These maps include up
to 18 points, at half-beam ($11''$) spacing. In the case of NGC~1333,
a 55-point map was made, encompassing the locations of both previous
\dammo\ detections. The integration time was 2 minutes per point,
using frequency switching with a frequency shift of 8~MHz.
Subsequently, deep integrations on the \nddh\ lines were performed at
the locations of the \nhhd\ peaks.

\subsection{Submillimeter-wave observations}
\label{sec:cso}

Submillimeter observations were carried out in 2002 and 2003 using the
345 GHz facility receiver and spectrometers of the Caltech
Submillimeter Observatory (CSO) on Mauna Kea (Hawaii). The data were
taken under average weather conditions (225 GHz zenith opacity
$\sim$0.1). The CSO FWHM beam size at 309 GHz is $\sim$25$''$, and the
main beam efficiency at the time of the observations, determined from
total-power observations of planets, was $\sim$60\%. Typical
calibration uncertainties are $\sim$25\%. The pointing of the
telescope was stable to within 5 arcsec. We used the 1.5 GHz, 500 MHz,
and 50 MHz bandwidth acousto-optical facility spectrometers. The
spectral resolution of the high-resolution, 1024 channel 50~MHz
spectrometer is $\sim$0.15~MHz or 0.15~km\,s$^{-1}$. All spectra were
taken in the double beam switched observing mode with a secondary
chopper throw of 180$''$.
 
\subsection{Centimeter-wave observations}
\label{sec:arecibo}

Observations of the \nddd\ inversion lines in the Barnard~1 cloud were
carried out between 2003 October 31 and November 11 at the Arecibo
Observatory in Puerto Rico, using the new L-Band Wide facility
receiver. Four subcorrelators of the interim spectrometer system were
used, each having 1024 lags. A high-resolution subcorrelator with a
total bandwidth of 0.781~MHz (0.144~\kms\ velocity resolution) was
centered on the frequency of the strongest \nddd\ hyperfine component,
1589.0178~MHz (Table~\ref{freq}) . Two additional subcorrelators with
bandwidths of 3.125~MHz, centered on the frequencies of 1588.4043 and
1588.6316~MHz, covered the remaining \nddd\ hyperfine components with
a velocity resolution of 0.576~\kms. In addition, a 1.563~MHz
bandwidth subcorrelator, centered on the frequency of the
1667.3590~MHz OH line, was used to monitor system performance during
the long integration. Quasars B02231+313 and B0340+048 were used for
pointing and calibration. The data were taken in position switching
mode with the reference position at a right ascension offset by 5
minutes of time. The typical system temperature referred to above the
earth's atmosphere was \about 25~K. The total on source integration
time was 756 min. The main beam plus first sidelobe efficiency of the
Arecibo Telescope is $\about 70\%$ at 1.6 GHz.

\section{Results}
\label{sec:res}

\subsection{Maps of \nhhd\ emission}
\label{sec:iram_maps}

\begin{figure}[ht]
\begin{center}
 \resizebox{\hsize}{!}{\includegraphics[angle=-90]{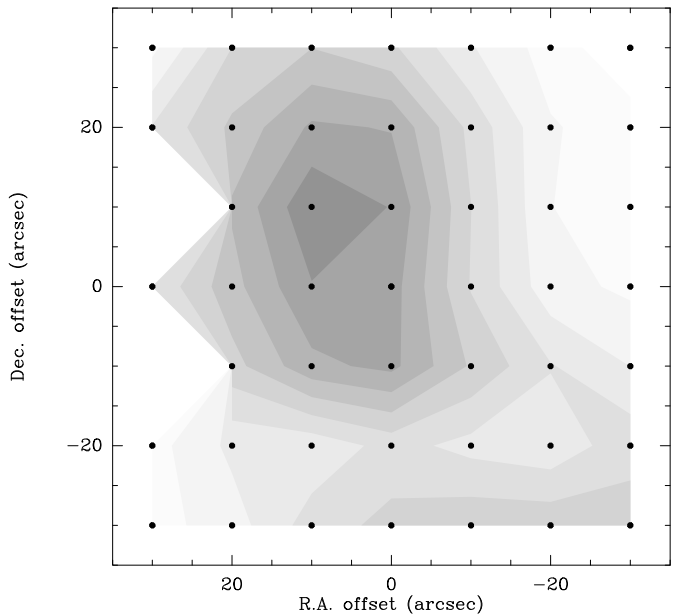}}
\caption{\label{fig:nh2d_maps}
  Map of $\int T_{\rm mb} dV$ of the main hyperfine component of the
  \nhhd\ $1_{11}-1_{01}$ (ortho) 85.9~GHz transition in Barnard~1.
 The coordinates of the (0,0) position are R.A.(J2000) = 03:33:20.8,
Dec.(J2000) = +31:07:34.8  .
  Greyscale steps are: 0.5 to 5.0 by 0.5 \kkms.}
\end{center}
\end{figure}

\begin{figure}[ht]
\begin{center}
\resizebox{\hsize}{!}{\includegraphics[angle=-90]{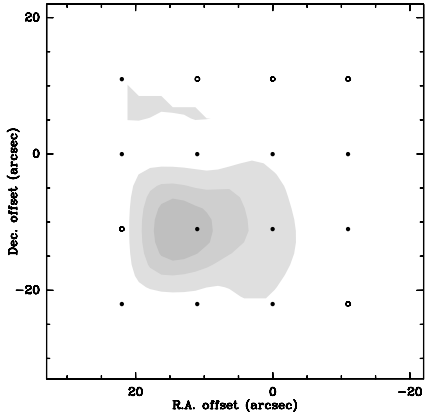}}
\caption{\label{fig:nh2d_map_hh1}Map of the integrated \nhhd\
  $1_{11}-1_{01}$ (para) 110.2~GHz emission in HH-1.
 The coordinates of the (0,0) position are R.A.(J2000) = 05:36:17.3,
Dec.(J2000) = -06:46:08.2  .
  Greyscale steps are 0.3 to 0.6 by 0.15 \kkms. Unfilled circles denote
  positions where no emission was detected.}
\end{center}
\end{figure}

\begin{figure}[ht]
\begin{center}
 \resizebox{\hsize}{!}{\includegraphics[angle=-90]{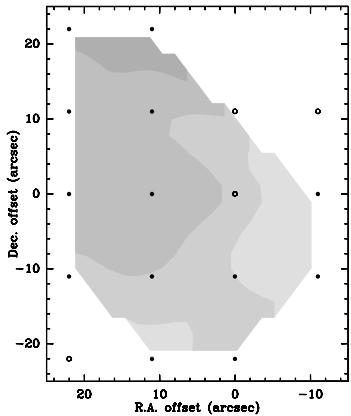}}
\caption{\label{fig:nh2d_map_omc2} As previous figure, for OMC-2.
 The coordinates of the (0,0) position are R.A.(J2000) = 05:35:26.3,
Dec.(J2000) = -05:09:49.4  .
  Greyscale steps are 0.1 to 0.5 by 0.1 \kkms.} 
\end{center}
\end{figure}

\begin{figure}[ht]
\begin{center}
 \resizebox{\hsize}{!}{\includegraphics[angle=-90]{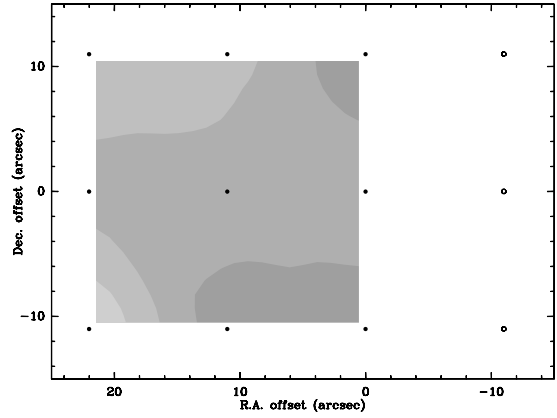}}
 \caption{\label{fig:nh2d_map_s68} As previous figure, for S68N.
 The coordinates of the (0,0) position are R.A.(J2000) = 18:29:47.5,
Dec.(J2000) = +01:16:51.4  .
  Greyscale steps are 0.1 to 0.7 by 0.1 \kkms.} 
\end{center}
\end{figure}

\begin{figure}[ht]
\begin{center}
\resizebox{\hsize}{!}{\includegraphics[angle=-90]{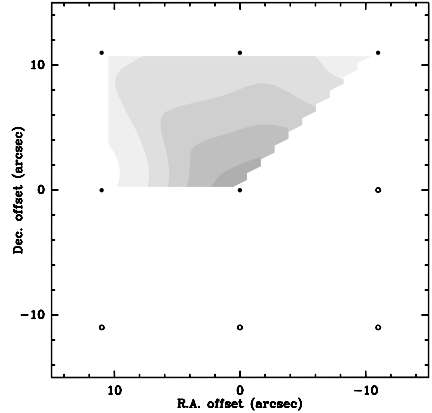}}
\caption{\label{fig:nh2d_map_2264}As previous figure, for NGC~2264C.
 The coordinates of the (0,0) position are R.A.(J2000) = 06:41:11.6,
Dec.(J2000) = +09:29:25.1  .
  Greyscale steps are 0.1 to 0.7 by 0.1 \kkms.} 
\end{center}
\end{figure}

\begin{figure}[ht]
\begin{center}
  \resizebox{\hsize}{!}{\includegraphics[angle=-90]{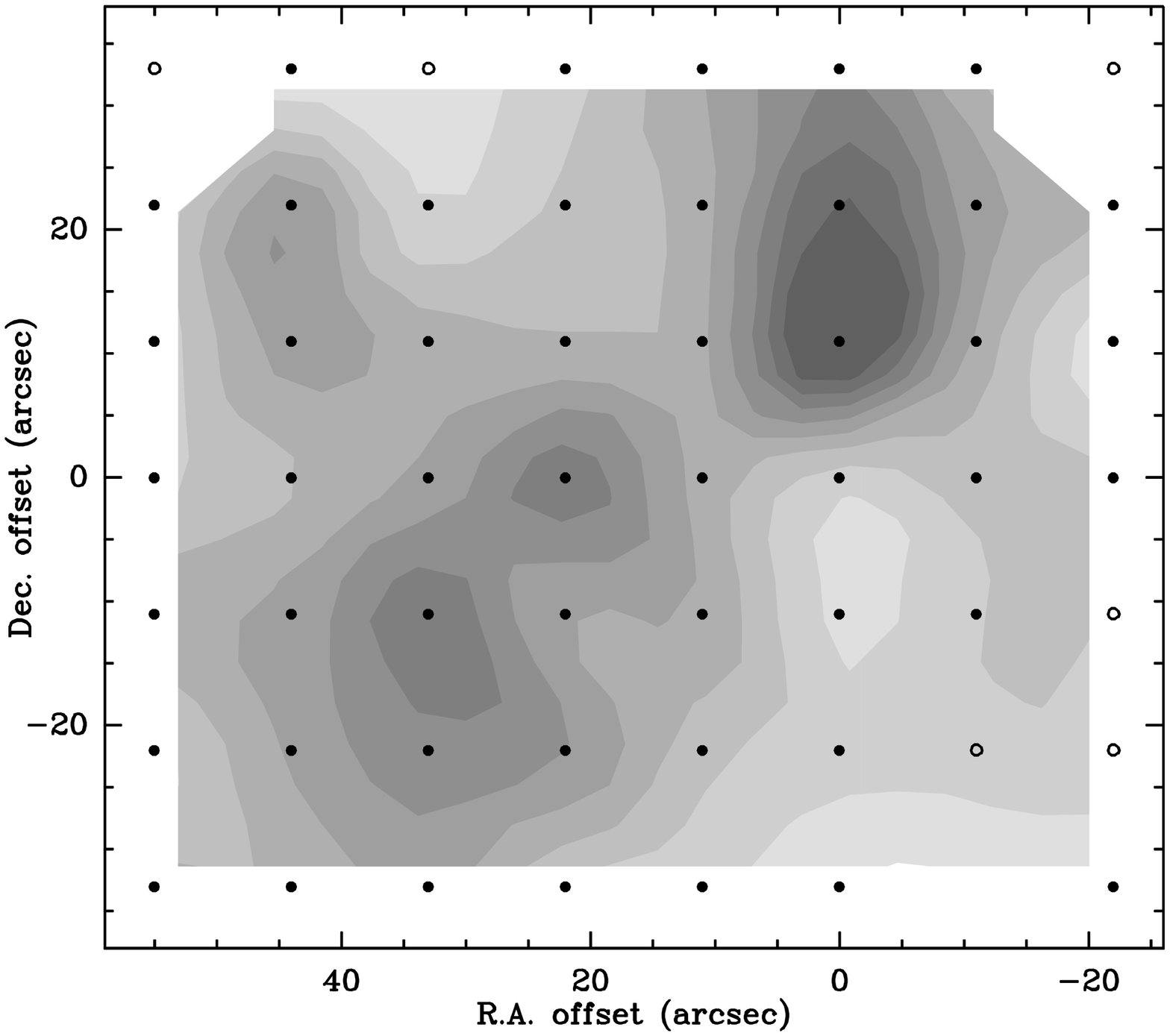}}
\end{center}
  \caption{\label{fig:1333} As previous figure, for NGC~1333.
 The coordinates of the (0,0) position are R.A.(J2000) = 03:29:10.3,
Dec.(J2000) = +31:13:32.2  .
  Greyscale steps are 0.1 to 0.9 by 0.1 \kkms.}
\end{figure}

Figures~\ref{fig:nh2d_maps} to~\ref{fig:1333} show the spatial
distribution of \nhhd\ emission as observed with the IRAM 30m. In
Barnard~1 and HH-1, the emission has a distinct peak at a slight
offset from our nominal central position given in Table \ref{tab:pos}.
On the other hand, the emission in OMC-2, S68N and NGC~2264C is not
strongly peaked. The map of NGC~1333 indicates that \nhhd\ does not
peak at the position of the protostar IRAS~4A at (0,0). Instead, the
emission peaks in two lobes, one located 10--20$''$~N of IRAS~4A and
the other to the W-SW of it, roughly between offsets (20,0) and
(30,--20). This distribution follows that of the DCO$^+$ 3$\to$2
emission, mapped at the CSO by \cite{lis04}, and is consistent with
the observation that the \dammo\ emission in NGC~1333 is stronger at
the offset position (23,--6) than at IRAS~4A \citep{vdtak02}.

Table~\ref{tab:nhhd} reports the column density of \nhhd\ at those
positions where 
the fitting of the hyperfine structure was successful, calculated
following \citet{tine00}. Emission of \nhhd\ is clearly detected in
other cases, as displayed in the maps, but no optical depth could be derived
as the individual spectra have a low signal to noise ratio.
The HFS method within the CLASS package
gives the total optical depth of the line and the radiation
temperature, assuming a beam filling factor of unity. Since this
latter assumption may not always be fulfilled, the table also gives
column density estimates for the cases \txc=5~K and 10~K. We
calculated the partition functions at these two considered
temperatures. The partition functions of \nhhd, \nddh, and \nddd\ for
an excitation temperature of 10~K are 9.37, 9.51, and 38.3,
respectively. The corresponding values for \txc=5~K are 4.47, 4.21,
and 17.23, respectively.

The absence of an \nhhd\ peak at NGC~1333 IRAS~4A could partly be due
to excitation effects. However, the $3_{22}\to3_{12}$ line of \nhhd\
is not detected at this position. The upper limits for this line in
NGC~1333, Barnard~1 and NGC~2264C~and G are \tmb$<$70--100~mK in
0.11~\kms\ channels. For S68N, OMC-2, HH-1 and CB~17, which were
observed during daytime, the corresponding upper limits are
\tmb$<$200-300~mK.

\begin{table*}[htbp]
  \caption{Observed \nhhd\ line parameters and derived column densities. 
    Numbers in parentheses denote uncertainties in units of the last decimal.}
  \label{tab:nhhd}
  \begin{center}
  \begin{tabular}{rrrrrrrrrr}
\hline \hline
\noalign{\smallskip}
Source    & offset & $V_0$ & $\Delta V$ & $T_R$ & $\tau$ & \txc\ & \multicolumn{3}{c}{$N^a$} \\
          &        &       &            &       &        &       & \multicolumn{3}{c}{\hrulefill} \\
          & arcsec & \kms  & \kms       & K     &        & K     & \multicolumn{3}{c}{$10^{14}$~\scm} \\
\noalign{\smallskip}
\hline
\noalign{\smallskip}

Barnard 1: \\
 +0  &  +0 & 6.78(1) &  0.76(2) &  1.3(2) &  4.0(5) &  4(3) &  5.8(7) &  4.3(5) &  2.9(3) \\
 +11 & +11 & 6.76(1) &  0.69(4) &  2.1(5) &  2.5(5) &  5(3) &  2.1(5) &  2.4(5) &  1.6(3) \\
 +11 &  +0 & 6.82(1) &  0.67(2) &  2.2(2) &  2.7(3) &  6(3) &  2.2(2) &  2.6(2) &  1.7(2) \\
 +11 &--11 & 6.88(1) &  0.72(3) &  1.9(4) &  2.6(5) &  5(3) &  2.6(5) &  2.7(5) &  1.8(3) \\
  +0 &--11 & 6.80(1) &  0.70(3) &  1.6(4) &  2.7(6) &  5(3) &  2.9(6) &  2.7(5) &  1.8(4) \\
--11 &--11 & 6.66(3) &  0.97(6) &  1.1(4) &  2.5(8) &  4(3) &  5(2)   &  3(1)   &  2.3(8) \\
--11 & +11 & 6.78(2) &  0.74(5) &  1.1(3) &  3.4(8) &  4(3) &  5(1)   &  3.6(9) &  2.4(6) \\
 +22 &  +0 & 6.83(2) &  0.73(6) &  2(1)   &  1.2(7) &  6(5) &  0.9(5) &  1.3(7) &  0.9(5) \\
 +22 & +11 & 6.72(2) &  0.61(5) &  2(1)   &  1.6(7) &  6(4) &  1.0(4) &  1.4(6) &  0.9(4) \\
 +22 &--11 & 6.90(2) &  0.70(6) &  1.2(5) &  2.4(9) &  4(3) &  3(1)   &  2.4(9) &  1.6(6) \\

 +0 &--22 &  9.21(4) &  0.47(7) &  0.8(6) &  3(2) &  4(4) &  4(3) &  2(1) &  1(1) \\

\noalign{\smallskip}
\hline
\noalign{\smallskip}

\end{tabular}
\end{center}

$^a$ Values in column 8 are for \txc\ equal to the estimate in column 7, 
     and in columns 9 and 10 for \txc=5~K and 10~K, respectively.
     Uncertainties in $N$ include only those arising from the uncertainties in $\tau$.
     
\end{table*}

\subsection{Spectra of \nddh\ emission}
\label{sec:iram_spectra}

\begin{figure}[ht]
\begin{center}
\resizebox{5cm}{!}{\includegraphics{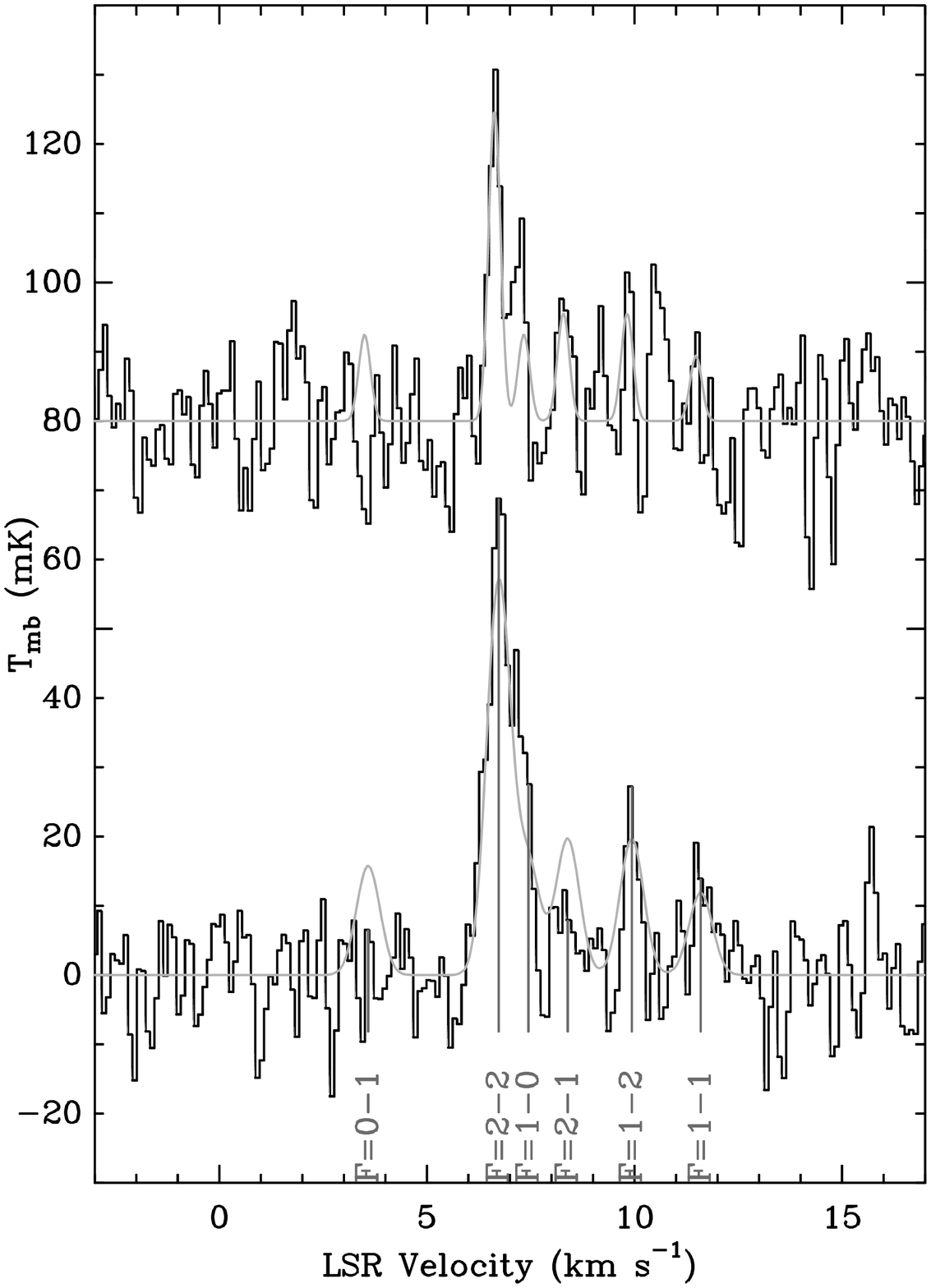}}
\caption{\label{fig:nddh}  
Spectra of the \nddh\ $1_{10}-1_{01}$ ortho (bottom) and para (top)
  transitions towards Barnard~1, with our fit to the hyperfine 
structure overplotted.}
\end{center}
\end{figure}

Emission in the $1_{10}\to1_{01}$ lines of \nddh\ was detected only in
Barnard~1; upper limits were obtained in other sources. The spectra
have rms levels of \tmb=5--11~mK in 0.11 \kms\ channels. The \nddh\
lines toward Barnard~1 are the strongest observed to date, and the
spectra (Fig.~\ref{fig:nddh}) are the first ones to show the hyperfine
components of the transition. 
The hyperfine structure has been calculated by L.~Coudert (priv.\ comm.)
and the corresponding frequencies and relative strengths are given in
Table~\ref{nddh:freq}.
Our fit to the hyperfine structure,
indicates a low optical depth ($\tau < 0.1$) for the lines.

\begin{table}[h] 
 \caption{Frequencies of the hyperfine components of the
 $1_{10} \rightarrow 1_{01}$ transitions of 
  para and ortho \nddh\ .}
  \label{nddh:freq}
  \begin{center}
    \leavevmode
    \begin{tabular}[h!]{ccc}
      \hline \hline \\[-5pt]
        Transition & $\nu ({\rm MHz})$ & Relative Strength \\[+5pt]
      \hline \\
     Ortho   &     &  \\
      \hline \\
       $F$=1--1   &  110811.1130 & 1/12 \\
       $F$=1--2   &  110811.7265 & 5/36 \\
       $F$=2--1   &  110812.2972 & 5/36 \\
       $F$=1--0   &  110812.6467 & 1/9  \\
       $F$=2--2   &  110812.9107 & 5/12 \\
       $F$=0--1   &  110814.0735 & 1/9  \\
      \hline\\[-0.0cm]
     Para   &     &  \\
      \hline \\
       $F$=1--1   &  110895.0298 & 1/12 \\
       $F$=1--2   &  110895.6433 & 5/36 \\
       $F$=2--1   &  110896.2108 & 5/36 \\
       $F$=1--0   &  110896.5635 & 1/9  \\
       $F$=2--2   &  110896.8243 & 5/12 \\
       $F$=0--1   &  110897.9823 & 1/9  \\
      \hline\\[-0.0cm]
      \end{tabular}
  \end{center}
 \end{table}

The column densities of \nddh\ have been calculated from the observed
line strengths, using spectroscopic constants from \citet{roueff00}.
We assume optically thin emission, but take deviations from the
Rayleigh-Jeans law into account. Table~\ref{tab:nddh} presents the
results. We assume \txc\ =10~K for positions close to young stellar
objects and 5~K for other positions. The numbers given in the table
supersede previous values given in \citet{waterloo}.

\begin{table*}[htbp]
  \caption{Observed \nddh\ line parameters (or 1$\sigma$ upper limits)
    and derived column densities. 
    Numbers in parentheses denote uncertainties in units of the last decimal.  
    Positions are the central positions given in Table \ref{tab:pos} unless
    an offset is indicated.}
  \label{tab:nddh}
  \begin{center}
  \begin{tabular}{llrrrrc}
\hline \hline
\noalign{\smallskip}
Position & ortho/ & $V_0$ & $\Delta V$ & $\int T_{\rm mb}dV$ & \txc\ & $N$(o+p) \\
         & para   & \kms   & \kms       & mK \kms   & K  & $10^{12}$~\scm \\
\noalign{\smallskip}
\hline
\noalign{\smallskip}

Barnard 1 & o     & 6.76(2) & 0.67(5) & 64(4) & 5.0 & 97(6) \\
          & p     & 6.97(2) & 0.34(5) & 30(4) & 5.0 & 75(10) \\
HH-1      & o     & ...     & 0.5$^b$ & $\le 2.7$ & 10.0& $\le 1.1$ \\
          & p     & ...     & 0.5$^b$ & $\le 2.2$ & 10.0& $\le 1.5$ \\
S68N      & o     & ...     & 0.8$^b$ & $\le 2.8$ & 10.0& $\le 1.1$ \\
          & p     & ...     & 0.8$^b$ & $\le 2.8$ & 10.0& $\le 1.9$ \\
NGC~1333: \\
(0,0)     & o     & ...     & 1.6$^a$ & $\le 3.6$ & 10.0& $\le 1.5$ \\
          & p     & ...     & 1.6$^a$ & $\le 3.6$ & 10.0& $\le 2.5$ \\
(0,11)    & o     & ...     & 1.6$^a$ & $\le 2.2$ & 10.0& $\le 0.9$ \\
          & p     & ...     & 1.6$^a$ & $\le 2.2$ & 10.0& $\le 1.5$ \\ 
(23,-6)   & o     & ...     & 1.0$^a$ & $\le 2.1$ & 5.0& $\le 3.2$ \\
          & p     & ...     & 1.0$^a$ & $\le 1.8$ & 5.0& $\le 4.5$ \\
(33,-22)  & o     & ...     & 1.0$^a$ & $\le 3.5$ & 5.0& $\le 5.3$ \\
          & p     & ...     & 1.0$^a$ & $\le 3.9$ & 5.0& $\le 9.7$ \\

\noalign{\smallskip}
\hline
\noalign{\smallskip}

  \end{tabular}
  \end{center}

$^a$ Value taken from \dammo\ \citep{vdtak02}

$^b$ Value taken from \nhhd\ (Table~\ref{tab:nhhd})

\end{table*}

Simultaneous observations of the \nddh\ $2_{20}-2_{11}$ line 
did not result in any detections. Upper limits to \tmb\ range from 11
to 22~mK in 0.11~\kms\ channels (\tsys\ of 400--500~K), except for
S68N and NGC~1333 (33,--22), where worse weather (\tsys\ $\sim$1000~K)
limited the sensitivity to 40--50~mK.

\subsection{Spectra and maps of \dammo\ emission}
\label{sec:nd3}

\begin{table*}[htbp]
  \caption{\label{tab:nd3} Observations of ND$_3$.  Offset positions from
  nominal central position are given in arc seconds, enclosed in parentheses.}
  \begin{center}
\begin{tabular}{lrrrrc}
\hline \hline
\noalign{\smallskip}
Position  & $V_0$ & $\Delta V$ & $\int T_{\rm mb}dV$ & \txc\ & $N$ \\
            & \kms   & \kms       & K \kms   & K  & $10^{12}$~\scm \\
\noalign{\smallskip}
\hline
NGC~1333 &  7.2	 & 1.6(6) &		0.071(21) & 10 & 0.29 (9)$^a$\\
NGC~1333 (23,-6)	&  6.5	&	1.0(2)	&	0.144(43) & 5 &
1.3(4) \\
NGC~1333 (-24,106) &  7.8 & 0.7(2) & 	0.078(17) & 5 & 0.70(15) \\
Barnard~1 (0,0)     & 6.5    & 0.82(7)  & 0.307(19)  & 5 & 2.8(7)$^b$\\ 
Barnard~1 (-31,113) &        & ...    & $\leq 0.045$  & 5 & $\leq 0.4$
\\
LDN~1544C	&  6.9	& 0.34(10) &	0.120(24) & 5 & 1.1(2) \\
LDN~1630 (20,35) &  ...	 & ...  &	$\leq 0.096$ & 5 & $\leq 0.9$ \\
LDN~134N	&  ...	& ... &	 $\leq 0.036$ & 5 & $\leq 0.32$ \\
LDN~1689N  (0,0) &   3.3 &  0.42(4) & 	0.165(17) & 5 & 1.49(15) \\
LDN~1689N (-10,-10) &  3.4 & 	0.44(3) & 0.202(23) & 5 & 1.8(2)\\
LDN~1689N (0,-20) &  3.3 & 	0.42(3) & 0.288(26) & 5 & 2.6(2)\\
LDN~483 &   5.3	 &  1.5(3) & 	0.081(19) & 5 & 0.7(2) \\
\noalign{\smallskip}
\hline
\end{tabular}
\end{center}
$^a$ van der Tak et al. (2002) ;
$^b$ Lis et al. (2002). The column density given here is given for
\txc = 5~K. 
\end{table*}

\begin{figure}[htbp]
  \begin{center}
  \resizebox{\hsize}{!}{\includegraphics{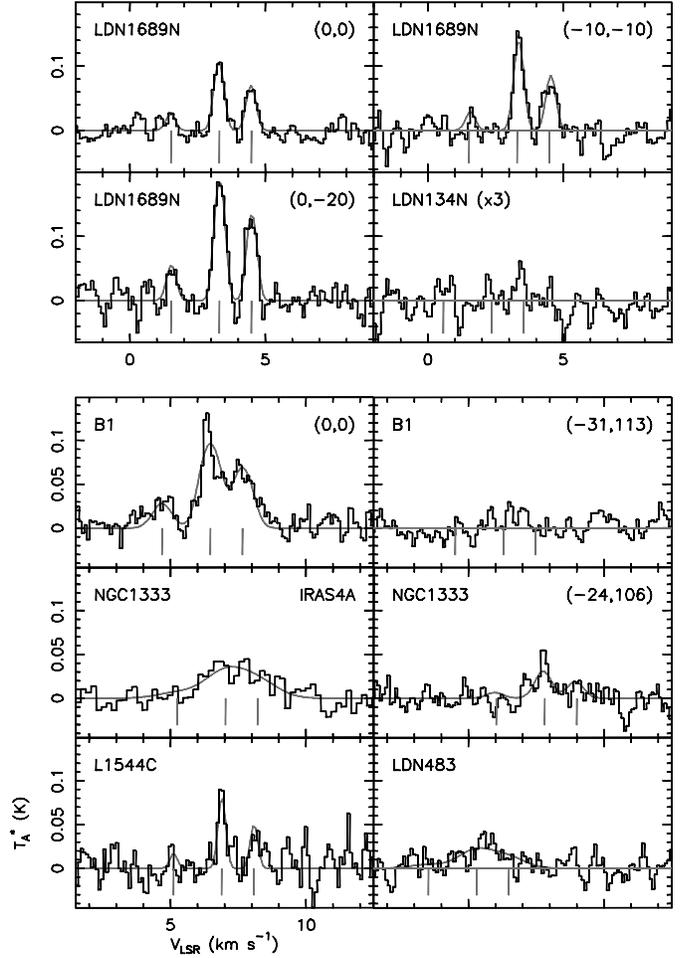}}  
  \end{center}
  \caption{Spectra of \dammo\ 309.9 GHz emission, observed with the
    CSO. The Barnard 1 (B1) spectrum is from Lis et al. (2002) and the
    NGC~1333 IRAS~4A spectrum from van der Tak et al. (2002).}
  \label{fig:nd3spec}
\end{figure}

\begin{figure}[htbp]
  \begin{center}
  \resizebox{\hsize}{!}{\includegraphics[angle=-90]{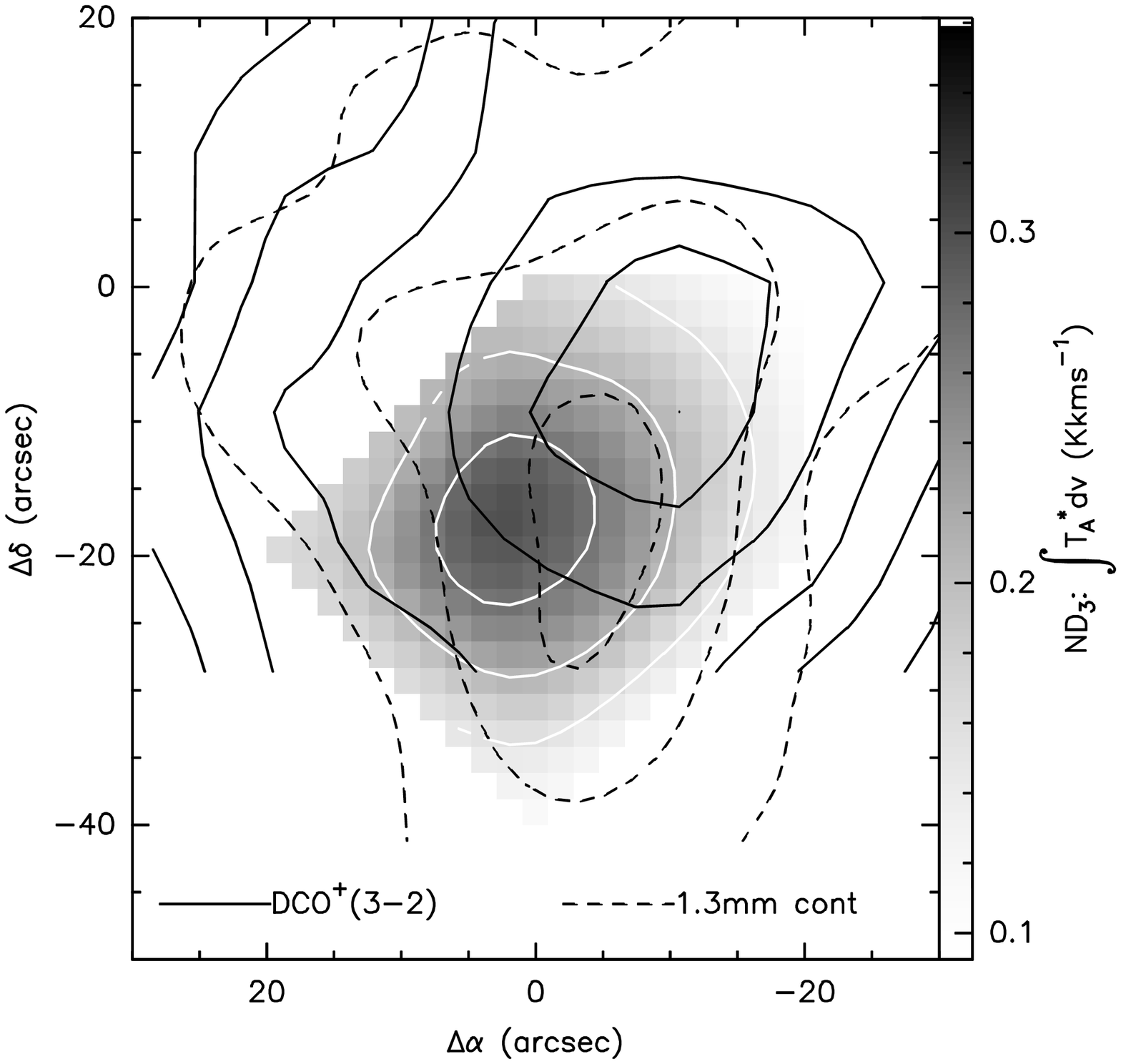}}  
  \end{center}
  \caption{Intensity of the \nddd\ emission in LDN~1689N integrated
 over velocities between 2.8 and 5~\kms, (greyscale and white
 contours), with overlayed contours of DCO$^+$(3-2) emission (solid
 black lines) and 1.3~mm dust continuum emission (dashed black lines).
 The coordinates of the (0,0) position are R.A.(J2000) = 16:32:29.5,
Dec.(J2000) = -24:28:52.6  .
 Contour levels are 50, 70, and 90\% of the peak for \nddd, 50 to 90\%
 of the peak, with an interval of 10\% for DCO$^+$, and 4, 6, and 8\%
 of the peak (toward IRAS~16293, outside the plot) for dust continuum.}
  \label{fig:nd3map}
\end{figure}

Table~\ref{tab:nd3} and Figure~\ref{fig:nd3spec} summarize the results
of the \dammo\ observations. In addition, Figure~\ref{fig:nd3map}
shows a map of \dammo\ in LDN~1689N, with overlaid contours of
DCO$^+$~(3--2) and 1.3~mm dust continuum emission. The emission is
seen not to peak on IRAS~16293, $\sim 90^{\prime\prime}$ to the east
of the map center, but rather at the outflow interaction region
between the YSO and a pre-stellar core. Many deuterated molecules
peak at this position, possibly because the outflows from IRAS~16293 A
and B, which are not co-eval, have desorbed grain mantles and
compressed the gas, which has subsequently cooled
\citep{lis02:l1689,stark04}. Indeed, the emission peaks at a velocity
offset from that of the core (Table~\ref{tab:pos}).

\subsection{ND$_3$ inversion lines}

\begin{figure}
\begin{center}
\includegraphics[angle=-90,width=8cm]{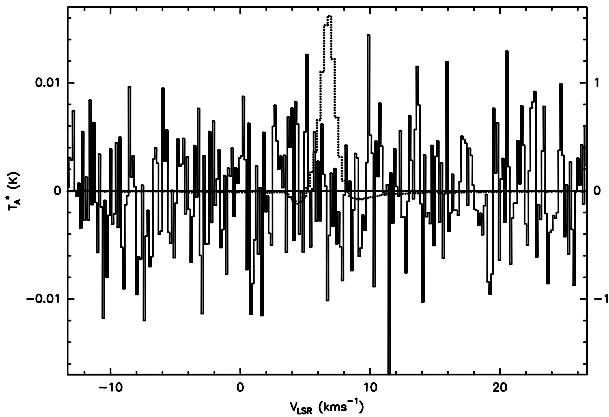}
\end{center}
\caption[f1.eps]{High-resolution spectrum at the frequency of the
  1589.0178~MHz \nddd\ hyperfine component in Barnard~1 (left scale).
  The dotted line shows the 1667.3590~MHz OH line at the same position
  (right scale).}
\label{arecibo}
\end{figure}

The high-resolution spectrum at the frequency of the strongest
hyperfine component of the 1.59~GHz \nddd\ inversion line in Barnard~1
is shown in Figure~\ref{arecibo}. No emission is detected with a $3
\sigma$ upper limit for the integrated intensity of 0.011~K\kms,
computed over a 2~\kms\ wide velocity interval and corrected for the
main beam efficiency. This upper limit has to be further corrected for
the source coupling to the 2\arcmper 5 Arecibo beam. The spatial
distribution of \nddd\ in the Barnard 1 cloud has not been determined.
Therefore to estimate the coupling efficiency, we assume that the
\nddd\ emission has the same spatial distribution as the \dcop~(3--2)
emission, as mapped with the CSO \citep{lis04}. The \dcop\
intensity in a 2.5\arcm\ beam is decreased by a factor of \about 2
compared to the peak value in the 30\arcs\ CSO beam. We thus estimate
a $3 \sigma$ upper limit for the integrated intensity of the
1589.0178~MHz \nddd\ inversion line in the CSO beam to be \about
0.022~K\kms.

To compute the corresponding upper limit for the \nddd\ column density
we use Eq.~(2) of \citet{lis02:nd3}, with $A = 5.23 \times
10^{-11}$~s$^{-1}$, $g_u = 24$, and $E_u = 11.97$~K, The \nddd\
partition function $Q=38.3$ for $T_{ex} = 10$~K, leading to a $3
\sigma$ upper limit of $3.0 \times 10^{13}$~\pscm\ for the \nddd\
column density, after including a factor of 2 correction for the
remaining hyperfine components (see Table~\ref{freq}). For $T_{ex} =
5$~K, the partition function $Q = 17.2$ and the corresponding $3
\sigma$ upper limit for the \nddd\ column density is $7.0 \times
10^{13}$~\pscm. These limits are an improvement over previous
Effelsberg results \citep{waterloo}, but still factors of $\sim$20
above the values derived from the rotational lines \citep{lis02:nd3}.

\begin{table}[ht]
  \caption{\nddd\ inversion frequencies within the $J_K=1_1$ level
derived from \cite{vanveld:02}.} 
  \label{freq}
  \begin{center}
    \leavevmode
    
    \begin{tabular}[ht!]{ccc}
      \hline \hline \\[-5pt]
        Transition & $\nu ({\rm MHz})$ & Relative Strength \\[+5pt]
      \hline \\[-5pt]
       $F$=0--1   &  1587.4645 & 1/9 \\
       $F$=2--1   &  1588.3969 & 5/36 \\
       $F$=1--1   &  1589.0114 & 1/12 \\
       $F$=2--2   &  1589.0086 & 5/12 \\
       $F$=1--2   &  1589.6231 & 5/36 \\
       $F$=1--0   &  1590.5545 & 1/9 \\
      \hline\\[-0.0cm]
      \end{tabular}
  \end{center}

\end{table}

\subsection{Summary of ammonia deuterium fractionation}

Table \ref{tab:summary} lists the column densities and fractionation
ratios for ammonia isotopologues in our sample. Only sources where
\dammo, or three other isotopologues have been detected are listed.
Care was taken that the data are in matching beams (20--40$''$)
whenever possible. For example, there are significant discrepancies in
the column density derivations from \citet{shah01} and
\citet{hatchell03} towards NGC~1333 IRAS~4A. The \ammo\ data for
NGC~1333, Barnard~1 and LDN~134N have been taken with the Effelsberg
telescope with a $\sim$40$''$ beam. Data for LDN~1544C and LDN~483
refer to a larger beam ( $\sim$100$''$) (see references listed in
Table \ref{tab:summary}), and the \ammo\ column density for LDN~1544
and LDN~483 may thus be underestimated.

Ammonia fractionation of the various isotopologues is seen to vary
from source to source, whereas grain surface chemistry predicts that
the deuterium fractionation should scale as atomic D/H ratio, and be
constant to within a factor of a few. However, we cannot rule out the
grain surface scenario, as the predictions are only qualitative
\citep{rodg01}. For most sources, [\nhhd]/[\ammo] and [\nddh]/[\nhhd]
ratios are of the same order of magnitude, $\sim 10 - 20$\%, while the
[\nddd]/[\nddh] ratio is significantly smaller ($\sim 2 - 5$\%). 
The assumption of one single excitation temperature for all
lines in one source may not be valid, since the lines have
very different excitation conditions (Table~\ref{obs:lines}). This
uncertainty may affect the column density ratios in Table~\ref{tab:summary}
because the column densities of some isotopologues are
derived from ground state lines and others from excited
states. 
From the observational side, the largest
uncertainty is thus in the radiative transfer modelling of the data.
Collisional cross sections for ammonia isotopologues are needed for a
more accurate analysis.

The \dammo\ to \ammo\ ratio varies between $\sim 1-2 \times 10^{-3}$
in LDN~1689N, Barnard 1, and the depleted core of LDN~1544, where
\hhdp\ and \ddhp\ have been detected (\citealt{caselli03} and
\citealt{vastel04}), to less than 1.6$\times$10$^{-4}$ for the dark
cloud LDN~134N. Such values represent prodigious enhancement factors
over the cosmic ratio of 12 orders of magnitude to less than 11 orders
of magnitude which deserve specific chemical studies. We present in
the next section a gas phase chemical model and the corresponding
detailed results.

\begin{table*}[htbp]
  \caption{\label{tab:summary} Summary of observations of ammonia fractionation}
  \begin{center}
  \begin{tabular}{lcccc|cccc}
\hline \hline
\noalign{\smallskip}
Source & NH$_3$   &  NH$_2$D       & ND$_2$H     & ND$_3$ & [\nhhd] / & [\nddh] / & [\dammo] / & [\dammo] / \\
 & 10$^{15}$~\scm & 10$^{14}$~\scm & 10$^{13}$~\scm & 10$^{12}$~\scm & [\ammo] & [\nhhd] & [\nddh] & [\ammo] \\
\noalign{\smallskip}
\hline
\noalign{\smallskip}
NGC~1333  & 1.4$\pm$0.4$^a$ & 3.9$\pm$0.8$^a$ & $\le$0.5$^h$ & 0.29$\pm$0.09$^d$ & 0.28$\pm$0.14 & $\le$0.01 & $\ge$0.06   & 2.1$\pm$1$\times$10$^{-4}$   \\
Barnard~1 & 2.5$\pm$0.2$^i$ & 5.8$\pm$0.7$^g$ & 8.6$\pm$0.8$^h$  &
2.8$\pm$0.7$^f$ & 0.23$\pm$0.05 & 
0.15$\pm$0.03 & 0.033$\pm$0.01 & 1.1$\pm$0.5 $\times$10$^{-3}$  \\
LDN~1544C  & 0.20$\pm$0.035$^k$ & 0.26$\pm$0.01$^k$ &  & 1.1 $\pm$0.2$^e$  & 0.13$\pm$0.03  &   &  & 5.5$\pm$1$\times$10$^{-3}$ \\
LDN~134N  & 2.0$\pm$0.2$^j$ & 2.0$\pm$0.4$^b$ & 1.0$\pm$0.2$^c$  & $\le$0.32$^e$ & 0.10$\pm$0.03 & 0.05$\pm$0.02 & $\le 0.032$ & $\le$1.6$\times$10$^{-4}$  \\
LDN~1689N &1.8$\pm$0.2$^k$ & 3.4$\pm$0.7$^l$ &7.6$\pm$1.1$^l$ &
1.8$\pm$0.2$^e$* & 0.19$\pm$0.05 & 0.22$\pm$0.05 & 0.024$\pm$0.01 &
1.0$\pm$0.5 $\times$10$^{-3}$ \\
LDN~483 & 1.4$\pm$0.1$^m$ &    &    & 0.7$\pm$0.2$^e$  &  &   &   &   5.0$\pm$1.2$\times$10$^{-4}$  \\
\hline
\end{tabular}
\end{center}

$^a$ \citet{hatchell03},   
$^b$ \citet{tine00},  
$^c$ \cite{roueff00}, 20 $\%$ accuracy assumed,
$^d$ \cite{vdtak02}, 
$^e$ Table~\ref{tab:nd3}, * the -10", -10" offset value has been given
as it corresponds to the \nddh\ peak observed by \citet{loinard01},
$^f$ \cite{lis02:nd3},
$^g$ Table~\ref{tab:nhhd}
$^h$ Table~\ref{tab:nddh},
$^i$ \cite{bachiller90}, 10 $\%$ accuracy assumed, 
$^j$ \citet{ungerechts80}, 10 $\%$ accuracy assumed, 
$^k$ \citet{shah01},10 $\%$ accuracy assumed,
$^l$ \citet{loinard01},
reanalysis of the para transition of \nhhd\ from the original data,
$^m$ \citet{jijina99}, \cite{anglada97}, 10$\%$ accuracy assumed.
\end{table*}

\section{A model of gas-phase deuterium chemistry}
\label{sec:models}

\subsection{Approach and general results}
\label{sec:model_setup}

We have updated the preliminary gas phase chemical network, which was
used to interpret our first detection of ND$_3$ towards Barnard~1, as presented
in \cite{lis02:nd3}. We have introduced the formation of \hh, HD, and
\dd\ on grain surfaces by following the simple approach presented in
\cite{leb95}. The role of the grains (passive,
although very significant) is to trap the condensable species, which
are then no longer available in the gas phase. The resulting chemical
abundances may differ considerably from the so-called typical values,
relevant for a standard molecular cloud such as TMC--1. Such effects are
labelled as depletion, and observations show a clear correlation of
deuterium fractionation and CO depletion in dense pre-stellar cores.
This trend results, to first order, from the shorter collision time
between dust grains and molecules when the density increases. 

In order
to mimic this effect, we consider three different models, in which the
density and depletion are varied. We consider a {}``standard'' molecular
cloud similar to TMC1 with a density of n(H$_2$) = 10$^4$ \ccm, a
temperature $T$ of 10~K and abundances relative to molecular hydrogen of
TMC--1, i.e. [C]/[H$_2$] = $7 \times 10^{-5}$; [O]/[H$_2$] = $2 \times
10^{-4}$; [N]/[H$_2$] = $2 \times 10^{-5}$. The relative abundance of
sulfur is not well known and is taken as [S]/[H$_2$] = $3.6 \times
10^{-7}$. In addition, we introduce a representative metal [M] which
does not contribute to the molecular complexity, and only undergoes
charge transfer and radiative recombination reactions, with
[M]/[H$_2$] = $3 \times 10^{-8}$ (low metal case). This model is
referred as model 1. For all models, we neglect possible deuterium depletion
on grains and have adopted a gas phase deuterium abundance of [D]/[H]
=$1.5 \times 10^{-5}$ as derived from the latest determinations
\citep{lin:03}. Models 2 and 3 have densities n(H$_2$) = 10$^5$ \ccm\
and 10$^6$ \ccm  and carbon and oxygen depletion factors of 
5 and 15, respectively. Nitrogen is kept constant in the
three models as ISO spectra have not indicated that ammonia ices are
abundant. The temperature has been kept equal to 10~K and the cosmic-ray
ionization rate is taken as 2 $\times$ 10$^{-17}$ s$^{-1}$.  
Table~\ref{tab:models} lists the predicted molecular abundances for the
three models. 

\begin{table*}[htbp]
  \caption{\label{tab:models} Prediction of fractional abundances 
 relative to molecular hydrogen}
  \begin{center}
  \begin{tabular}{llll}
\hline \hline
\noalign{\smallskip}
  & Model 1 & Model 2 & Model 3 \\
\hline 
\noalign{\smallskip}
n(H$_2$) (cm$^{-3}$) & 10$^4$ &  10$^5$ & 10$^6$\\
Depletion  & 1 & 5 & 15 \\
x(e$^-$)$^a$ & $6.05 \times 10^{-8}$ &  $3.74 \times 10^{-8}$  & $2.91 \times 10^{-8}$ \\
\hline
H  & $2.25 \times 10^{-4}$ &  $2.26 \times 10^{-5}$  & $2.27 \times 10^{-6}$ \\
HD  & $2.75 \times 10^{-5}$ &  $2.65 \times 10^{-5}$  & $2.56 \times 10^{-5}$ \\
D  & $1.22 \times 10^{-6}$ &  $4.43 \times 10^{-7}$  & $8.12 \times 10^{-8}$ \\
D$_2$  & $5.84 \times 10^{-7}$ &  $1.49 \times 10^{-6}$  & $2.18 \times 10^{-6}$ \\
CO & $6.79 \times 10^{-5}$ &  $ 1.33 \times 10^{-5}$  & $4.47 \times 10^{-6}$ \\
C & $1.72 \times 10^{-6}$ &  $ 5.64 \times 10^{-7}$  & $1.64 \times 10^{-7}$ \\
H$_3^+$ & $1.03 \times 10^{-8}$ &  $ 3.01 \times 10^{-9}$  & $4.78 \times 10^{-10}$ \\
H$_2$D$^+$ & $5.80 \times 10^{-10}$ &  $ 4.45 \times 10^{-10}$  & $1.05 \times 10^{-10}$ \\
D$_2$H$^+$ & $3.13 \times 10^{-11}$ &  $ 6.47 \times 10^{-11}$  & $2.33 \times 10^{-11}$ \\
D$_3^+$ & $1.29 \times 10^{-12}$ &  $ 8.35 \times 10^{-12}$  & $5.20 \times 10^{-12}$ \\
\hline 
NH & $4.89 \times 10^{-9}$ &  $ 6.80 \times 10^{-9}$  & $1.66 \times 10^{-9}$ \\
ND & $1.18 \times 10^{-9}$ &  $ 4.67 \times 10^{-9}$  & $1.70 \times 10^{-9}$ \\
NH$_2$ & $6.01 \times 10^{-8}$ &  $ 8.48 \times 10^{-8}$  & $1.25 \times 10^{-8}$ \\
NHD & $2.99 \times 10^{-9}$ &  $ 7.64 \times 10^{-9}$  & $1.51 \times 10^{-9}$ \\
ND$_2$ & $3.05 \times 10^{-11}$ &  $ 2.05 \times 10^{-10}$  & $5.09 \times 10^{-11}$ \\
NH$_3$ & $5.49 \times 10^{-8}$ &  $ 8.82 \times 10^{-8}$  & $2.00 \times 10^{-8}$ \\
NH$_2$D & $2.71 \times 10^{-9}$ &  $ 8.80 \times 10^{-9}$  & $2.94 \times 10^{-9}$ \\
ND$_2$H & $8.82 \times 10^{-11}$ &  $ 6.10 \times 10^{-10}$  & $2.37 \times 10^{-10}$ \\
ND$_3$ & $3.44 \times 10^{-12}$ &  $ 5.14 \times 10^{-11}$  & $2.14 \times 10^{-11}$ \\
HCN & $7.13 \times 10^{-8}$ &  $ 3.78 \times 10^{-8}$  & $7.92 \times 10^{-9}$ \\
DCN & $2.40 \times 10^{-9}$ &  $ 3.15 \times 10^{-9}$  & $8.90 \times 10^{-10}$ \\
HNC & $3.29 \times 10^{-8}$ &  $ 1.95 \times 10^{-8}$  & $4.02 \times 10^{-9}$ \\
DNC & $1.54 \times 10^{-9}$ &  $ 2.35 \times 10^{-9}$  & $6.28 \times 10^{-10}$ \\
HCO$^+$ & $9.67 \times 10^{-9}$ &  $ 1.02 \times 10^{-9}$  & $8.01 \times 10^{-11}$ \\
DCO$^+$ & $ 2.85 \times 10^{-10}$ &  $ 6.85 \times 10^{-11}$  & $8.15 \times 10^{-12}$ \\
N$_2$H$^+$ & $2.66 \times 10^{-10}$ &  $ 5.28 \times 10^{-10}$  & $1.68 \times 10^{-10}$ \\
N$_2$D$^+$ & $9.24 \times 10^{-12}$ &  $ 4.25 \times 10^{-11}$  & $1.80 \times 10^{-11}$ \\
\hline
\end{tabular}
\end{center}
$^a$ fractional ionization.
\end{table*}

We consider 210 gas phase species, including \hh, H$_2$O, CH$_4$, 
\ammo, H$_2$S, H$_2$CS, all deuterated variants thereof, and
corresponding ions. A total of 3000 chemical reactions are included
in the chemical network. In addition to the reactions of \hhhp,
CH$_3^+$, C$_2$H$_2^+$ with HD, which are driving the deuteration as
already introduced in \cite{watson74}, we have considered the
subsequent deuteration of molecular hydrogen by introducing the
reactions of \hhdp\ and \ddhp\ with HD, D, and D$_2$ as studied
experimentally at low temperatures by \cite{gers02}. Some of these reactions
have also been introduced by \cite{rob04} and by \cite{wal04} 
for the completely depleted
case. 
Table~\ref{tab:chi} gives the fractionation reactions, the 
rate coefficients of the forward corresponding reaction and the 
endothermicities involved in the reverse reactions. These values
are obtained from the ground level energies without differentiating
between ortho and para forms of the various species involved.

\begin{table}[htbp]
  \caption{\label{tab:chi} Fractionation reactions involving \hhhp,
   CH$_3^+$ and C$_2$H$_2^+$ with HD, D$_2$ and D.}
  \begin{center}
  \begin{tabular}{llll}
\hline \hline
\noalign{\smallskip}
  & \ \ \ \ k                 & $\Delta E$ & Ref. \\
  & cm$^3$ s$^{-1}$ & \ K & Ref. \\
\hline
\noalign{\smallskip}
 \hhhp\  + HD  $\rightleftharpoons$ \hhdp\ + \hh\ & 3.5(-10) & 232 & 1 \\
 \hhdp\  + HD  $\rightleftharpoons$ \ddhp\ + \hh\ & 2.6(-10) & 187 & 1 \\
 \hhhp\  + \dd\  $\rightleftharpoons$ \hhdp\ + HD & 1.75(-10) & 152 & 1 \\
 \hhhp\  + \dd\  $\rightleftharpoons$ \ddhp\ + \hh\ & 1.75(-10) & 340 & 1 \\
 \ddhp\  + HD  $\rightleftharpoons$ \dddp\ + \hh\ & 2.00(-10) & 234 & 1 \\
 \hhdp\  + \dd\  $\rightleftharpoons$ \ddhp\ + HD & 3.50(-10) & 109 & 1 \\
 \ddhp\  + \dd\  $\rightleftharpoons$ \dddp\ + HD & 3.50(-10) & 100 & 2 \\
 \hhhp\  + D  $\rightleftharpoons$ \hhdp\ + H & 1.0(-9) & 632 & 1 \\
 \hhdp\  + D  $\rightleftharpoons$ \ddhp\ + H & 1.0(-9) & 600 & 1 \\
 \ddhp\  + D  $\rightleftharpoons$ \dddp\ + H & 1.0(-9) & 600 & 4 \\

CH$_3^+$ +  HD $\rightleftharpoons$ CH$_2$D$^+$ + \hh\ & 2.6(-10) & 375 & 1\\
CH$_3^+$ +  \dd\ $\rightleftharpoons$ CD$_2$H$^+$ + \hh\ & 4.4(-10) & 375 & 5\\
CH$_3^+$ +  \dd\ $\rightleftharpoons$ CH$_2$D$^+$ + HD
 & 6.6(-10) & 375 & 1\\
CH$_2$D$^+$ +  HD $\rightleftharpoons$ CD$_2$H$^+$ + \hh\ & 1.0(-9) & 370 & 1\\
CH$_2$D$^+$ +  \dd\ $\rightleftharpoons$ CD$_3$$^+$ + \hh\ & 3.0(-10) & 370 & 3\\
CH$_2$D$^+$ +  \dd\ $\rightleftharpoons$ CD$_2$H$^+$ + HD & 9.0(-10) & 370 & 3\\
CD$_2$H$^+$ + HD $\rightleftharpoons$ CD$_3$$^+$ + \hh\ & 5.64(-10) & 370 & 3\\
CD$_2$H$^+$ +  \dd\ $\rightleftharpoons$ CD$_3$$^+$ + HD & 4.5(-10) & 370 & 3\\

C$_2$H$_2^+$ +  HD $\rightleftharpoons$ C$_2$HD$^+$ + \hh\ & 7.5(-10) & 550 & 1\\
C$_2$HD$^+$ +  HD $\rightleftharpoons$ C$_2$D$_2^+$ + \hh\ & 7.5(-10) & 300 & 1\\
C$_2$H$_2^+$ +  \dd\ $\rightleftharpoons$ C$_2$HD$^+$ + HD & 7.0(-10) & 550 & 2\\
C$_2$HD$^+$ +  \dd\ $\rightleftharpoons$ C$_2$D$_2^+$ + HD & 7.0(-10) & 550 & 2\\
C$_2$H$_2^+$ +  D $\rightleftharpoons$ C$_2$HD$^+$ + H & 7.5(-10) & 250 & 2\\
C$_2$HD$^+$ +  D $\rightleftharpoons$ C$_2$D$_2^+$ + H & 7.5(-10) & 150 & 2\\

\hline \hline
\end{tabular}
\end{center}

References: 1. \citet{gerlich02}; 2. estimate; 3. \citet{smith82};
4. \citet{wal04}; 5. \citet{anicich03}
\end{table}

As already discussed in \cite{lis02:nd3} and \cite{rou04}, we have
taken special care regarding the dissociative recombination reactions
and introduced several updates such as the new dissociative
recombination rate coefficient of H$_3$$^+$ \citep{Mcc03}, the new
branching ratios of the dissociative recombination of \nnhp\ by
\citet{Geppert04}, who found that NH is a major product channel and
the recent measurement of ND$_4^+$ \citep{oje04}. In addition, we have
generalized the result of recent studies of dissociative recombination
reactions of partially deuterated molecular ions (H$_2$D$^+$, HDO$^+$
and HD$_2$O$^+$) which show that the release of H atoms is favoured
compared to that of deuterium atoms. We assume that the branching
ratios differ by a factor of 2 between H and D ejection, following the
recommendations of M. Jensen (private communication)\footnote{In
\cite{lis02:nd3} we had assumed a factor 3 to 1}. The influence of
this assumption is discussed in Section~\ref{sec:appl}
 
We can derive the following general trends from these models:
\begin{enumerate}
\item The fractional ionization decreases as the density increases, as
   expected. The main source of electrons is provided by the
   representative metal which may only recombine or charge transfer
   and does not contribute to the chemical complexity.
\item The deuterium reservoir is provided by HD. However, as the
   density increases, molecular D$_2$ becomes a significant deuterium
   containing species. It is then important to account properly for
   the gas phase reactions involving doubly deuterated hydrogen.
\item Model 2 presents the highest molecular fractional abundances for the
  N-bearing species, while HCN, HNC, and HCO$^+$ follow the same
  pattern as C and CO and have higher fractional abundances in model 1. 
\item The gas phase chemistry changes significantly between models 2
   and 3. Indeed, with the high depletion assumed for model 3, the
   \hhhp\ chemistry is driven by reactions with HD rather than
   reactions with the abundant neutrals O, CO, etc... This results in
   significant variations in the fractional abundances of saturated
   species such as \ammo, HCN, HNC, and HCO$^+$ and N$_2$H$^+$
   molecular ions.
\item D$_2$ is the sole species, which has higher abundance in model
  3, as compared to model 2.
\item The combination of high density and depletion results in very
   efficient production of deuterated species. The abundances of the
   multiply deuterated species such as \dddp\ and \dammo, are enhanced
   by between 10 and more than 12 orders of magnitude relative to
   those expected from the cosmic deuterium abundance, in the various
   models we have considered.
\item All models predict high fractional abundances of ammonia progenitors
   such as NH, ND, NH$_2$, NHD and ND$_2$.

\end{enumerate}

\subsection{Comparison with the observations}
\label{sec:appl}

We now focus on the comparison of the deuterium fractionation values
of the various species with the different observed values. 
Table~\ref{tab:compar} reports the deuterium fractionation ratio obtained
with the three previously described models and the observed values
towards a sample of clouds.
We also report the fractionation ratios of nitrogen containing molecules 
calculated when statistical approximation is assumed for the       
reaction products 
coming from dissociative recombination reactions, when the latter have 
not been studied
in the laboratory.  The resulting fractionation ratios differ at most 
by a factor of 2 in the nitrogen containing species.

\begin{table*}[htbp]
  \caption{\label{tab:compar} Comparison between predicted and observed
deuterium
fractionation}
  \begin{center}
  \begin{tabular}{l|lll|llll}
\hline \hline
\noalign{\smallskip}
  & Model 1 & Model 2 & Model 3  & LDN~134N  &  LDN~1689N  &  Barnard~1  &  LDN~1544\\
\hline
\noalign{\smallskip}
D:H & 0.0054 & 0.020 & 0.031  &   &    &     &   \\
D$_2$:HD & 0.022 & 0.056 & 0.083 &   &   &    &    \\
H$_2$D$^+$ : H$_3^+$ & 0.056 & 0.15 & 0.22 &   &    &    &  \\
D$_2$H$^+$ : H$_2$D$^+$ & 0.054 & 0.14 & 0.22 &   & 0.75$^a$  &   &   \\
D$_3^+$ : D$_2$H$^+$ & 0.043 & 0.15 & 0.27 &   &    &    &  \\
D$_3^+$ : H$_3^+$ & $1.32 \times 10^{-4}$ & $3.15 \times 10^{-4}$ & 0.013 &   &    &    &  \\ 
NH$_2$D : NH$_3$ & 0.049 & 0.10 & 0.15 & 0.1$^b$  & 0.19$^b$ & 0.23$^b$ & 0.13$^b$\\
NH$_2$D : NH$_3$$^*$ & 0.025 & 0.052 & 0.077 &   &  &  &  \\
ND$_2$H : NH$_2$D & 0.032 & 0.069 & 0.080& 0.05$^b$ & 0.22$^b$  & 0.15$^b$ &  \\
ND$_2$H : NH$_2$D$^*$ & 0.021 & 0.040 & 0.049&  &   &  &  \\
ND$_3$ : ND$_2$H & 0.039 & 0.084 & 0.090& $\le 0.032$ & 0.024$^b$& 0.033$^b$ & \\
ND$_3$ : ND$_2$H$^*$ & 0.035 & 0.075 & 0.078&  &   &  &  \\
ND$_3$ : NH$_3$ & $6.3 \times 10^{-5}$ & $5.8 \times 10^{-4}$ & $1.1  \times 10
^{-3}$& $\le 1.6 \times 10^{-4}$ & $1.0 \times 10^{-3}$$^b$ & $1.1 \times 10^{-3}$$^b$ & $5.4 \times 10^{-3}$$^b$ \\
DCN : HCN & 0.034 & 0.083 & 0.11  & 0.06$^c$ & 0.11$^e$ &   &  \\
DCN : HCN$^*$ & 0.019 & 0.049 & 0.075 &  &  &   &  \\
DNC : HNC & 0.047 & 0.12 & 0.16 & 0.12$^f$ & 0.091$^f$ &  &0.050$^g$\\
DNC : HNC$^*$ & 0.029 & 0.071 & 0.091&  &  &  &  \\
DCO$^+$ : HCO$^+$ & 0.030 & 0.067 & 0.10& 0.18$^d$   &  0.08$^e$ &     &   0.04$^h$\\
N$_2$D$^+$ : N$_2$H$^+$ & 0.035 & 0.087 & 0.11& 0.35$^d$  &    & 0.15$^i$   &  0.2$^h$ \\
\hline
\end{tabular}
\end{center}
$^*$  models employing a statistical approximation for the reaction products in
dissociative recombination reactions of non studied deuterated molecular ions\\
$^a$ \citet{vastel04}, para-\ddhp\ / ortho-\hhdp\   
$^b$ Table~\ref{tab:summary},
$^c$ \citet{turn01}, based on N(HN$^{13}$C) $\times$60,
$^d$ \citet{tine00}, 
$^e$ \citet{lis02:l1689}, based on N(H$^{13}$CN) and N(H$^{13}$CO$^+$) $\times$60,
$^f$ \citet{hiro01}, based on N(HN$^{13}$C) $\times$60,
$^g$ \citet{hiro03}, based on N(HN$^{13}$C) $\times$60,
$^h$ \citet{caselli02},
$^i$ \citet{gerin01}
\end{table*}

Following previous remarks, the deuterium fractionation ratio is seen
to increase from model 1 to model 3. The \ddhp\ / \hhdp\ ratio found
in LDN~1689N involves significantly excited levels (the upper level
energies of \hhdp\ and \ddhp\ involved are 104 and 83~K, respectively)
and may not reflect the actual value of the full \ddhp\ / \hhdp\ ratio
calculated in our models. \cite{wal04} and \cite{fl:04}
have considered a completely depleted chemistry with specific para and
ortho species and find a p-\ddhp\ / o-\hhdp\ consistent with the
observations.
 
However, the ensemble of data for this pre-stellar core (sometimes
also called IRAS~16293E) is spectacular. The level of deuteration is
particularly high. \citet{stark04} deduce a depletion of CO of the
order of 5 compared to TMC-1 (from [C$^{18}$O] / [H$_2$]= $3 \times
10^{-8}$ and [CO] : [C$^{18}$O] = 500 : 1), equal to the value assumed
in model 2, and a temperature in the range 12 $\leq$ T $\leq$ 16~K,
whereas we have performed calculations at 10~K. Comparison between
model 2 and observations toward LDN~1689N shows an agreement within a
factor of 2 for the reported species except for ND$_2$H which displays
a spectacular enrichment. Another source of deuteration than the one
assumed may be at work. A plausible mechanism is evaporation from dust
grains, resulting from the shocks produced by the outflows, as
suggested by \citet{lis02:l1689}. This hypothesis is also suggested by
the detection of doubly deuterated formaldehyde by \citet{cec:02} in
the same pre-stellar core.

The results displayed for LDN~134N deserve special comments as the
search of \dammo\ has been very deep and a very low upper limit has
been obtained. The density and depletion conditions are probably
between those taken for model~2 (\citealt{tine00,roueff00}) and
model~3 (\citealt{pa:05}). Comparison between models and observed
deuterium fractionation ratios is satisfactory except for \dammo\
where predictions are above the observed upper limit, and for
N$_2$H$^+$ which exhibits an impressive deuterium fractionation ratio
towards this line of sight. Such values may be obtained within the
present treatment if the dissociative recombination rate coefficient
of the deuterated N$_2$D$^+$ were significantly smaller than that of
N$_2$H$^+$, a possibility which is not excluded (such effects have
been measured for \hhhp\ and \dddp\ but not studied experimentally for
these nitrogen bearing ions).
 
\begin{figure}[htbp]
  \begin{center}
   \resizebox{7.5cm}{!}{\includegraphics{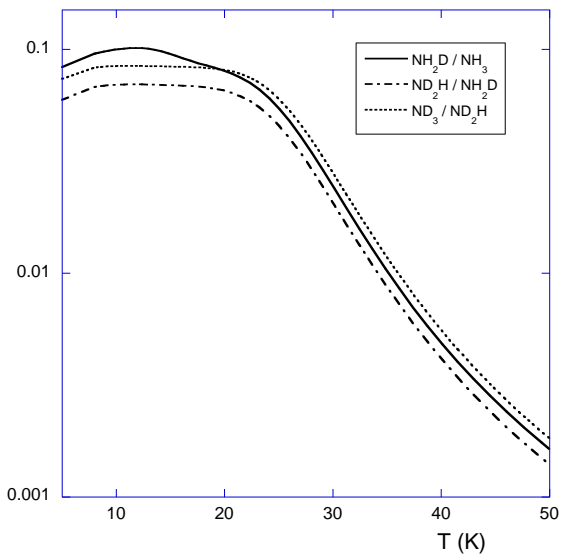}}
  \end{center}
  \caption{Results of model calculations of the ratio of different
  deuterated isotopologues of ammonia as a function of temperature.
  The conditions of model 2 have been employed. The full line shows
  \nhhd\ relative to \ammo\, the dotted line shows \nddh\ relative to
  \nhhd\, and the dash-dotted line shows \nddd\ reltive to \nddh\ .
  Note that \nddh\ relative to \nhhd\ is the lowest ratio for all the
  temperatures considered in this calculation. }
  \label{fig:mdl}
\end{figure}

\begin{figure}[htbp]
  \begin{center}
   \resizebox{7.5cm}{!}{\includegraphics{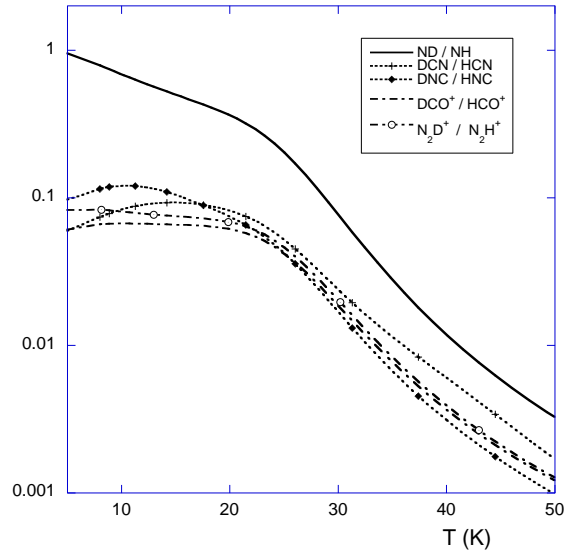}}
  \end{center}
  \caption{Results of model calculations of the ratio of different
  deuterated isotopologues of nitrogen containing molecules as a
  function of temperature. The conditions of model 2 have been
  employed. The full line displays ND/NH , crosses and diamonds on the
  dotted line shows DCN/HCN and DNC/ HNC respectively whereas the
  dashed-dotted line without and with open circles refer to
  DCO$^+$/HCO$^+$ and N$_2$D$^+$ and N$_2$H$^+$ respectively. }
  \label{fig:dcn}
\end{figure}

The observations in the Barnard 1 molecular cloud are reproduced
within a factor of 2 by Model 2 except for \nddh\ which is
underpredicted by a factor of 3. As temperatures in prestellar clouds
can vary between values as low as 7~K in the center and $\simeq$50~K
at large radii, we display in Figures~\ref{fig:mdl} and \ref{fig:dcn}
the relative abundances of deuterated isotopologues of ammonia and of
other molecules at temperatures between 5 and 50~K for the conditions
pertaining to model 2, with a molecular hydrogen density of 10$^5$
\ccm. The deuterium fractionation of the various molecules is of the
order of 10\% for temperatures between 5 and 20~K (except for ND/NH
where the value is about unity at low temperatures), and decreases
quickly for higher temperatures.

The peak in the deuterium fractionation of \nddd\ relative to \ammo\
at $T\approx 12$~K is directly linked to the barrier of the reaction
of N$^+$ with HD which is of the order of 16~K \citep{mar88}. 
Indeed, \cite{jacq90} measured lower ammonia fractionation ratio in
hot cores $\leq 0.01$ than in the dense cold cores studied here. The
temperature dependence is discussed by \cite{shah01}. The ammonia
fractionation remains close to 10\% for a kinetic temperature below 
30~K, but drops to a few percent in warmer regions. The turnover in the
model presented here occurs at a slightly lower temperature (22 K),
but the overall shape is fairly well reproduced, given the uncertainty
in the observed temperature measurements.

Another interesting feature is seen in Figure~\ref{fig:dcn}, where the
modelled DCN/HCN ratio is found to be smaller than that of DNC/HNC at
temperatures below 20~K, whereas the opposite is found at higher
temperatures. This trend is clearly found towards LDN~134N which is
known to be a cold source. Such a tendency had already been emphasized
by \citet{rg03} for a sample of dark cold sources. We have been led to
carefully consider the various channels leading to deuterated HCN and
HNC which are produced, in such gas phase chemistry, via the
dissociative recombination of HCND$^+$, DCNH$^+$ and HDNC$^+$.
Table~\ref{tab:reac} displays the values assumed.

\begin{table}[h]
\caption{\label{tab:reac}Assumed reaction rate coefficients and branching 
ratios in dissociative recombination reactions of deuterated 
HCNH$^+$ and H$_2$NC$^+$}
\begin{center}
\centering
\begin{tabular}{llll}
\hline    \hline

Reaction &   &    &  k (in 10$^{-7}$ cm$^{3}$s$^{-1}$) \\
\hline
 HCND$^+$ + electron & $\rightarrow$ & HCN + D & 1.16 (T/300)$^{-{0.5}}$\\
          &         &  DNC + H  &   2.33 (T/300)$^{-{0.5}}$  \\
 DCNH$^+$ + electron & $\rightarrow$ & DCN + H & 2.33 (T/300)$^{-{0.5}}$  \\
          &         &  HNC + D  & 1.16 (T/300)$^{-{0.5}}$  \\
 HDNC$^+$ + electron & $\rightarrow$ & HNC + D  &  0.58 (T/300)$^{-{0.5}}$  \\
          &         &  DNC + H  & 1.16 (T/300)$^{-{0.5}}$    \\
          &         &  NHD + C  & 1.75 (T/300)$^{-{0.5}}$    \\
\hline
\end{tabular}
\end{center}
\end{table}

We note that this prediction is not derived in the accretion models of
\citet{rob04}. DCN/HCN and DNC/HNC ratios become closer in LDN~1689N
where the temperature may be around 16~K as discussed by \citet{mi:90}
and \citet{stark04}, in nice agreement with the model predictions. In
addition, we display the temperature dependence of the deuterium
fractionation ratios of the HCO$^+$ and N$_2$H$^+$ molecular ions,
which are very similar.

The \dammo\ to \ammo\ ratio is highest for LDN~1544. However, the ammonia
column density in this source may be underestimated because of the
large beam size of the observations. High fractionations are also
obtained for Barnard~1 and LDN~1689N, for which we could benefit from
higher quality ammonia data. We therefore believe that the correction for the
ammonia column density is at most a factor of three and the \dammo\ to
\ammo\ ratio in LDN~1544 is close to the values obtained in Barnard~1 and
LDN~1689N. This line of sight is well studied and known
to be a highly depleted core \citep{caselli03}. The observations and
the results of model 3 agree within a factor of 3. \citet{rob04}
have made detailed modelling of LDN~1544 by introducing the density and
temperature structure of this prestellar core in a model including
accretion processes on dust grains and an extensive chemical network.
The overall predicted deuterium fractionation is larger by about one
order of magnitude than the present observations. However, a nice
agreement between models and observations is obtained for doubly
deuterated formaldehyde by \citet{rob04}.

Extremely young protostars have been found in the Barnard~1 molecular
cloud \citep{hirano99} and the depletion factor is about 3 compared to
TMC-1 (cf.\ \citealt{lis02:nd3}). Recent observations have revealed
the presence of HDS, D$_2$S \citep{vastel:03}, HDCS and D$_2$CS
\citep{mar:05} in this dark cloud, which deserves a specific chemical
analysis. The level of deuteration is quite high and probably
involves a source of deuteration in addition to that provided by gas
phase chemistry alone.

\section{Conclusions}
\label{sec:con}
The present study reports detection of fully deuterated ammonia
towards 5 lines of sight, with derived ND$_3$/NH$_3$ ratios between 2
$\times 10^{-4}$ and 1--2 $\times 10^{-3}$ in various pre-stellar
cores. A very low upper limit for this ratio (1.6 $\times 10^{-4}$)
has been found towards the dense dark cloud LDN~134N. These
observations imply a tremendous enhancement factor over the cosmic 
deuterium/hydrogen 
value which varies from about 11 to over 12 orders of magnitude. The
spatial extension is limited and the ND$_3$ peak does not coincide
with the DCO$^+$ peak in LDN~1689N, a source where such a study is
possible thanks to its exceptionally high deuterium enhancement.
Deuterated species do not peak towards protostars themselves, but at
offset positions, suggesting that protostellar activity decreases
deuteration built in the pre-stellar phase. We have studied the origin
of the deuterium enrichment in terms of a steady state gas phase
chemical model, which includes a limited number of molecules, but which
allows full deuteration of basic important intermediates. The role of
grains is simulated by introducing different values of the depletion
of C and O so that the important deuterated molecular ions H$_2$D$^+$,
D$_2$H$^+$, D$_3$$^+$, CH$_2$D$^+$, CD$_2$H$^+$, CD$_3$$^+$ are not
destroyed by reactions with O, CO, ... in the highly depleted regions.
Our models make the following predictions:

\begin{itemize}

\item Higher density models with higher C and O depletions yield
greater enhancement of the relative abundances of the deuterated
ammonia isotopologues.

\item The models predict abundance ratios of the deuterated ammonia
isotopologues, which agree reasonably well with new and existing
observations.

\item The fractionation ratios of ammonia deuterated isotopologues
remain large for temperatures as high as 20~K.

\item The temperature dependence of the ratios DCN/HCN and DNC/HNC is
not identical. Whereas DNC/HNC is larger at low temperatures, the
opposite becomes true at higher temperatures.

\item A very high deuterium fractionation is predicted for ND. We
also find a significant abundance of NHD in our models. Both species
have rotational transitions in the submillimeter domain and are
potentially detectable from the ground.

\item D$_2$ may be an important component of the cloud and the
corresponding reactions have to be introduced. There is,
unfortunately, no hope of detecting it under the conditions which
characterize cold, dark clouds.

\end{itemize}

\begin{acknowledgement}
  The authors thank the staffs of the Arecibo, CSO and IRAM 30m
  telescopes for their support. IRAM is an international institute for
  research in millimeter--wave astronomy, co--funded by the Centre
  National de la Recherche Scientifique (France), the Max Planck
  Gesellschaft (Germany) and the Instituto Geografico Nacional
  (Spain). The CSO is funded by the U.S.\ NSF through contract
  AST~22-09008. The Arecibo Observatory is part of the National
  Astronomy and Ionosphere Center, which is operated by Cornell
  University under a cooperative agreement with the National Science
  Foundation. We thank Laurent Coudert who calculated for us the hyperfine 
  structure
 of the various deuterated isotopologues of ammonia. 
 MG and ER acknowledge travel support from the CNRS/INSU research program PCMI.
\end{acknowledgement}

\bibliographystyle{aa}

\end{document}